\documentclass[
 reprint,
 superscriptaddress,
showpacs,preprintnumbers,
 amsmath,amssymb,
 aps,
prl
]{revtex4-1}

\usepackage{amsmath,amssymb,bbm,epsfig}
\usepackage{hyperref}

\usepackage{color}
\usepackage[usenames,dvipsnames]{xcolor}
\usepackage[abs]{overpic}

\newcommand{\be}{\begin{equation}}
\newcommand{\ee}{\end{equation}}
\newcommand{\ba}{\begin{array}}
\newcommand{\ea}{\end{array}}
\newcommand{\bqa}{\begin{eqnarray}}
\newcommand{\eqa}{\end{eqnarray}}

\newcommand{\um}{\mathbbm{1}}

\newcommand{\ket}[1]{\ensuremath{| #1 \rangle}}

\newcommand{\ketbra}[2]{\ensuremath{|#1 \rangle\langle #2|}}

\newcommand{\Htg}{H_\text{tg}}
\newcommand{\Omegatg}{\Omega_\text{tg}}
\newcommand{\Utg}{U_\text{tg}}
\newcommand{\Heff}{H_\text{eff}}
\newcommand{\Omegaeff}{\Omega_\text{eff}}
\newcommand{\Ueff}{U_\text{eff}}
\newcommand{\Ufl}{\tilde U}
\newcommand{\Hc}{\text{H.c.}}
\newcommand{\Nmax}{N}

\begin{document}

\title{Optimal Control of Effective Hamiltonians}

\author{Albert Verdeny}
\affiliation{Freiburg Institute for Advanced Studies, Albert-Ludwigs-Universit\"at, Albertstrasse 19, 79104 Freiburg, Germany}

\author{\L ukasz Rudnicki}
\affiliation{Freiburg Institute for Advanced Studies, Albert-Ludwigs-Universit\"at, Albertstrasse 19, 79104 Freiburg, Germany}
\affiliation{Center for Theoretical Physics, Polish Academy of Sciences,
Aleja Lotnik\' ow 32/46, 02-668 Warsaw, Poland}

\author{Cord A. M\"uller}
\affiliation{Centre for Quantum Technologies, National University of Singapore, Singapore 117543, Singapore}
\affiliation{Department of Physics, University of Konstanz, 78457 Konstanz, Germany}

\author{Florian Mintert}
\affiliation{Freiburg Institute for Advanced Studies, Albert-Ludwigs-Universit\"at, Albertstrasse 19, 79104 Freiburg, Germany}
\affiliation{Department of Physics, Imperial College London, London SW7 2AZ, United Kingdom}

\begin{abstract}
We present a systematic scheme for optimization of quantum simulations. Specifically, we show how polychromatic driving can be used to significantly improve the driving of Raman transitions in the Lambda system, which opens new possibilities for controlled driven-induced effective dynamics.
\end{abstract}

\date{\today}

\maketitle

In the past few years, one of the most active and promising research fields has been the design of quantum simulators, i.e. engineered controllable quantum systems utilized to mimic the dynamics of other systems. With this, new insight is expected to be gained in a variety of phenomena like high-temperature fractional quantum Hall states \cite{Tang11}, (non-)abelian gauge fields \cite{Hauke12,Banerjee13} and even relativistic effects \cite{Goldman09,FV13}.

Driven systems provide a powerful tool to simulate desired effective dynamics.
An important example  are laser-assisted Raman transitions between different electronic states and/or localized states of trapped atoms, which is a central pillar in a large number of quantum simulations. Because the direct coupling between low-lying energy states via dipole transitions is often forbidden by selection rules, an intermediate auxiliary state with higher energy is usually used to mediate the coupling. This so-called Lambda system is then specifically configured to imprint phases required to realize various spin-orbit couplings \cite{Cheuk12,Wang12} or to simulate the effect of gauge fields \cite{Miyake13,Aidelsburger13}.
Other prominent examples of driving-induced effective dynamics include shaken lattices \cite{Eckardt05,Lignier07,Struck12}, lattices with modulated interactions \cite{Rapp12} or driven graphene \cite{Iadecola13}.

Even though driven systems provide a powerful approach to perform quantum simulations, they often rely on approximations that currently situate them still far from the ideal quantum simulator.
For instance, in Raman transitions via a three-level Lambda system, the driving pulse produces an undesired population of the excited state.
These deviations between the desired and simulated dynamics accumulate during the evolution and become considerable after a sufficiently long time.
From the experimental side, however, spectacular progress has been made in the manipulation and control of quantum systems \cite{Braun13,Struck13}, so that accurate theoretical tools to choose the proper driving are necessary.
The field of optimal control theory \cite{Alessandro07,Krotov96} aims at such precise manipulation but, so far, it has primarily targeted properties at single instances in time \cite{Khaneja05,Reich13,Bartels13} whereas we are rather concerned with the behavior of a system during a continuous time window.

In this Letter, we provide a general systematic approach to improve quantum simulations by using pulse shaping techniques of optimal control theory.
We discuss in detail the optimal control of the Lambda system and rigorously show how an appropriately chosen polychromatic driving can significantly improve Raman transitions.
As a result, we do not only provide a proof of principle for the optimal control of effective Hamiltonians but also optimize a building-block used in a large variety of quantum simulations.

Consider the target dynamics $U_{\rm tg}=e^{-iH_{\rm tg}t}$, generated
by the target Hamiltonian $\Htg$ that we wish to simulate using a time-periodic 
driving Hamiltonian $H(t)=H(t+T)$. Its 
time--evolution operator
$U(t)=\mathcal{T}\exp[-i\int_{0}^{t}H(t')dt']$  then 
admits the Floquet decomposition \cite{Floquet83}
\begin{equation}
U(t)=\Ufl(t) \Ueff (t).\label{decomp}
\end{equation}
Here $\Ufl(t)$ is a $T$-periodic unitary satisfying $\Ufl(0)=\um$, $\Ueff(t)=e^{-i\Heff t}$
and $\Heff$ is a time-independent effective Hamiltonian  defined via
$U(T)=e^{-i\Heff T}$. 
$\Ufl$ describes fluctuations around 
the envelope evolution $\Ueff$.
These fluctuations become negligible if typical matrix elements $\Omegaeff$ of $\Heff$
are sufficiently small compared to 
the driving frequency $\omega=2\pi/T$. In this case the low-energy or long-time
dynamics of the periodically driven system is well described by
$\Heff$, which in turn should be chosen to match the target
Hamiltonian $\Htg$ to be simulated. 
Suppose now that the driving $H(t)$ contains a set  $\{f_n\}$ of control parameters. 
Our aim is to tune these parameters such that
the dynamics $U$ resembles the target dynamics $U_{\rm tg}$ as well as
possible. Different choices of $\{f_n\}$ can result in 
similar effective dynamics, but produce different fluctuations. In order to ensure the optimal simulation of a
given target Hamiltonian 
$H_{\rm tg}$ with least fluctuations, our scheme therefore consists in:
 
(i) Identifying the dependence of the effective Hamiltonian $\Heff$ 
on the control parameters $\{f_n\}$. Typically this can only
be achieved in an approximate manner,  where $\Heff = \sum_{k=0}^r
\Heff^{(k)}+O(\Omega_{\rm eff}\epsilon^{r+1})$ is known 
up to order $r$ of the small parameter $\epsilon =
\Omegaeff/\omega \ll 1 $.   

(ii) Constraining $\{f_n\}$ so
that $\Htg=\Heff$ to the same order $r$. 

(iii) Minimizing the target functional 
\begin{eqnarray}\label{targetF}
\mathcal{F}&=&\dfrac{1}{T}\int_{0}^{T} \| U(t)-\Utg(t)\|^{2}dt,  
\end{eqnarray}
where $||\cdot||^2={\rm \textrm{Tr}\left(\cdot^{\dagger}\cdot\right)}$
is the Hilbert--Schmidt norm,
under the constrained control parameters allowed by (ii).       
$\mathcal{F}$ shall also be approximated to the order $r$ consistent with (i-ii).   

In order to calculate $\Heff$, $U(t)$, and $\mathcal{F}$, we use the Magnus expansion
\cite{Magnus54,Blanes09} and write $U(t)=e^{-iM(t)}$ as
the exponential of a time-dependent operator 
$M(t)=\sum_{k=1}^\infty M_k(t)$. The first two terms of this series
are  $M_1(t) = \int_0^t H(t_1)dt_1$ and 
 \begin{eqnarray}
M_2(t)&=&-\dfrac{i}{2}\int_0^t dt_1 \int_0^{t_1}dt_2 [H(t_1),H(t_2)]. 
\end{eqnarray}
The $k$th-order Magnus operator $M_k(t)$ contains $k$-fold time integrals of $k-1$ nested commutators.
When $H(t)$ is $T$-periodic,  $M_k(T)$ is exactly of order $\epsilon^k$.
Thus, $\Heff^{(k)}=M_{k+1}(T)/T$, since the factor $1/T$ reduces the order of expansion by one. 
In this manner, $\Heff$, $U(t)$, and consequently $\mathcal{F}$ are determined up to order $r$, and our scheme (i-iii) yields parameters $\{f_n\}$ that ensure
the optimal simulation of  $\Htg$.

In the following, we exemplify the method described above with a case study of the 
degenerate Lambda system where $\ket{1} $ and $\ket{2}$ denote the two ground
states and $\ket{3}$ the excited state. The target Hamiltonian 
\begin{equation} \label{Htg}
\Htg = - \Omegatg \left(\ketbra{1}{2}+ \ketbra{2}{1}\right) 
\end{equation}  
generates Raman transitions within the ground-state manifold at a
rate $\Omegatg$ (the overall sign is chosen for later convenience). Our aim is to simulate this dynamics by driving the 
transitions $\ket{1}\leftrightarrow\ket{3}$ and $\ket{2}\leftrightarrow\ket{3}$
 with a suitably modulated Rabi frequency.
In the interaction picture, where the dynamics induced by the static Hamiltonian is absorbed in the state vectors, the driving Hamiltonian takes the form 
\begin{eqnarray}\label{noRWA}
H(t)&=& f(t)\left(1+e^{-in_0 \Delta t}\right)\left(\ketbra{1}{3}+\ketbra{2}{3}\right)+\Hc.
\end{eqnarray}
Importantly, we assume that the driving pulse 
\begin{equation}
f(t)=\sum_{n=1}^{\Nmax}f_{n}e^{-in\omega t}\label{series}
\end{equation}
is written as a general Fourier series in terms of $\omega$, the fundamental frequency of driving. 
Since we do not want the optimization to rely on strong intensity and
fast frequencies, the maximal frequency in the above pulse is the
detuning $\Delta=\epsilon_3-\omega_d=N\omega$ between the driving carrier and excitation frequency.   
In other words, no
Fourier components with frequency larger than $\Delta$ need to be
generated. $H(t)$ defined in Eq.~\eqref{noRWA} should be periodic with period
$T=2\pi/\omega$, which is the case if $\Delta$ is a
fraction of twice the driving carrier frequency, such that $n_0$ is a
integer. In the rotating wave approximation, one would neglect the
counter-rotating contribution $e^{-in_0 \Delta t}$ in \eqref{noRWA}; we
consider the general case and keep this term.

Let us first discuss the simplest example of a monochromatic (MC)
driving $f(t)=f_1 e^{-i\Delta t}$ at constant Rabi frequency by taking $N=1$ in eq.~\eqref{series}.  
Since the Magnus operator $M_{2l}(t)$ of even order contains even products of $H(t)$,
the corresponding effective Hamiltonian $H_{\rm eff}^{(2l-1)}$ has the desired structure of $\Htg$ with matrix
elements that couple
the ground states
\footnote{These terms also generate light-shift
  displacements of the excited and ground states which do not influence the target
dynamics between the two ground states.}.
Choosing  
\begin{equation}\label{f1MC}
|f_1|^2=\Omegatg\Delta \frac{1+n_0}{2+n_0},
\end{equation} 
the constraint $\Heff^{(1)}=\Htg$ can 
be fulfilled to first order since $\Heff^{(0)}=0$.  
The second-order term $H_{\rm eff}^{(2)}$, on the other hand, will generate undesired
transitions to the upper level via cubic powers of $H(t)$. 
With $f_1$ already fixed, one cannot impose $\Heff^{(2)}=0$, so that one always ends
up with an unwanted population in the excited state---except in the ideal limit of very strong, far-detuned  driving $|f_1|, \Delta \rightarrow \infty$ at fixed 
$\Omegatg$. Thus, with only one frequency,
one can neither accurately realize the desired unitary ground-state
dynamics nor simultaneously
minimize the fluctuations. 

Let us then take advantage of the general pulse \eqref{series} and implement the first constraint $\Heff^{(1)} = \Htg$ with
\begin{eqnarray}
\Omegatg &=&\dfrac{1}{\omega} \sum_{n=1}^{\Nmax} \dfrac{|f_n|^2}{\tilde{n}(1)},  \label{constr1} 
\end{eqnarray}
where $\tilde{n}(p)^{-1}\equiv n^{-p}+(n+\Nmax n_0)^{-p}$. 
[In the rotating wave approximation $n_0\to\infty$ this simplifies to $\tilde{n}(p)  = n^p$.]
Pushing the Magnus expansion to third order, we can now require 
$\Heff^{(2)}=0$ through the second constraint 
\begin{eqnarray}
0&=&\sum_{n=1}^{\Nmax} \dfrac{f_n}{\tilde{n}(1)}. \label{constr2}
\end{eqnarray}
The target functional to be minimized reads
\begin{equation}
\mathcal{F}^{\left(2\right)}=\frac{4}{\omega^{2}}  \sum_{n=1}^{\Nmax} \dfrac{|f_n|^2}{\tilde{n}(2)}. 
\label{target}
\end{equation}
We can solve the optimization problem now analytically
by introducing two Lagrange multipliers
$\lambda_1\in\mathbbm{R},\lambda_2\in\mathbbm{C}$ for the two
constraints \eqref{constr1} and \eqref{constr2}, respectively. The
optimal pulse parameters are found to be  
\begin{eqnarray}\label{secondnoRWA}
f_n&=&\omega \lambda_2\left(
  \dfrac{\tilde{n}(1)}{\tilde{n}(2)}-\lambda_1 \right)^{-1}.  
\end{eqnarray}
The Lagrange multipliers are determined by 
inserting this solution into
the constraints 
\eqref{constr1} and \eqref{constr2}. 
Dividing eq.~\eqref{secondnoRWA} by $\lambda_2$ shows that
$f_n/\lambda_2$ is real, and thus all $f_n$ as well as $\lambda_2$ can
be taken real.
Using Eqs.~\eqref{constr1} and \eqref{constr2}, the target functional (\ref{target}) can be rewritten as
$\mathcal{F}^{(2)}=4\lambda_1\Omegatg/\omega$,
such that the 
global minimum of the fluctuations is found with the minimal root
$\lambda_1$ of Eq.~\eqref{constr2} with \eqref{secondnoRWA} inserted. 
One should be able to suppress fluctuations more efficiently using more frequencies,  and indeed,
as $\Nmax\to\infty$ the minimal $\lambda_1$ tends to $\tilde{\Nmax}(1)/\tilde{\Nmax}(2)\sim 1/\Nmax$, such that $\mathcal{F}^{(2)}\rightarrow 0$.

With relatively little effort, one can take the calculation one step
further. Using the first 4 terms of the Magnus expansion, the
constraint \eqref{constr1}  can be
extended to third order $H^{(1)}_{\rm eff}+H^{(3)}_{\rm eff}=H_{\rm
  tg}$ with the constraint
\begin{eqnarray}
\Omegatg&=&\omega \left( A_1+2 B_3-4 A_1A_2 \right)\label{l2}
\end{eqnarray}
defined in terms of $A_p=\sum|f_n/\omega|^2/\tilde{n}(p)$  and $B_p=\sum
|f_n/\omega|^4/\tilde{n}(p)$.
Eq.~\eqref{target} does not change to this order since
$\mathcal{F}^{(3)}=0$. 
The optimal Fourier components $\{ f_n\}$ can still be chosen real and solve the coupled system of equations ($n=1,...,\Nmax$)
\begin{eqnarray}\label{analiticthird}
\frac{f_n}{\omega}\left[\dfrac{
    1+ 4A_1  \lambda_1}{\tilde{n}(2)}
-\lambda_1\dfrac{1- 4A_2}{\tilde{n}(1)}\right]  -
\dfrac{4\lambda_1f_n^3}{\tilde{n}(3)\omega^3} 
=\dfrac{\lambda_2}{\tilde{n}(1)}. 
\end{eqnarray}

\begin{figure}
(a) \includegraphics[scale=0.28]{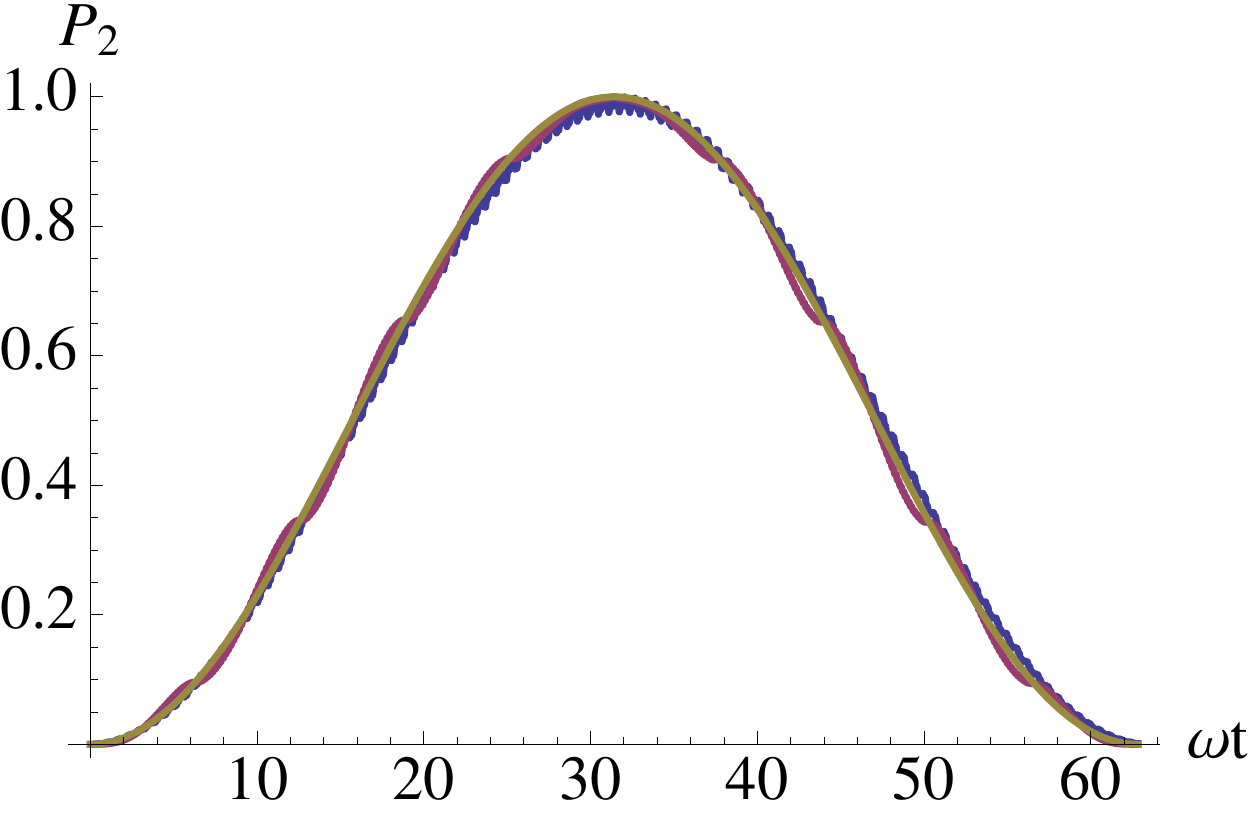}
(b) \includegraphics[scale=0.28]{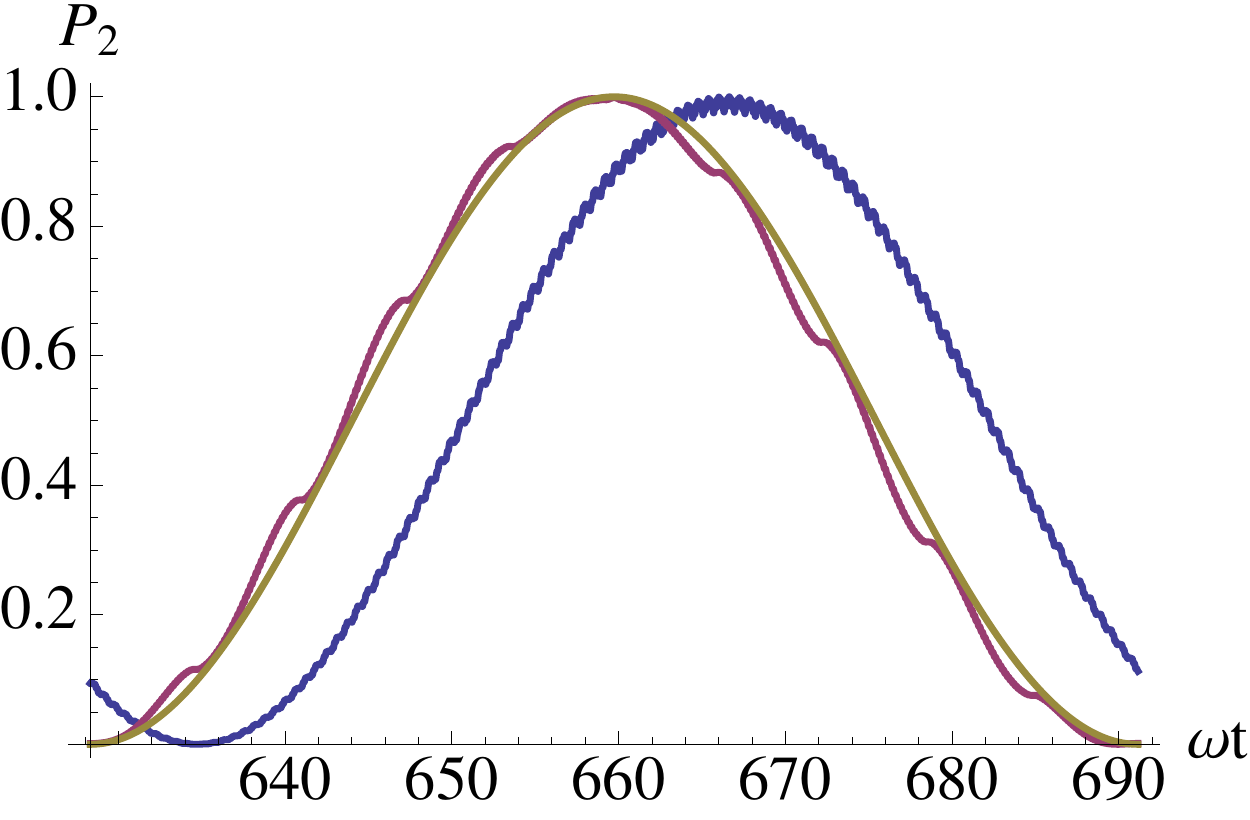}

(c) \includegraphics[scale=0.46]{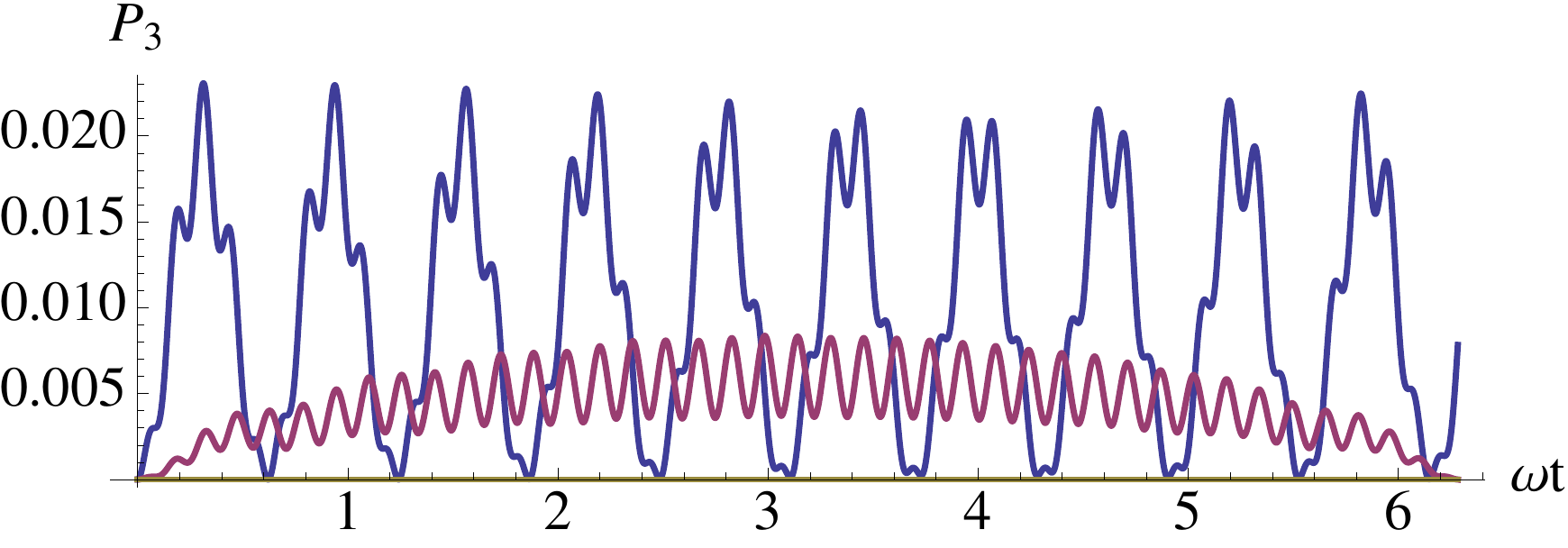}
\caption{\label{dynamics}(color online). (a) and (b) Transition probability $P_2(t)=|\langle
  2|U(t)|1\rangle|^2$ for the target dynamics
  (yellow), MC dynamics (blue) and third-order optimal dynamics with
  $N=10$ frequency components
  (red). (c) Population of the excited state $P_3=|\langle
  3|U(t)|1\rangle|^2$ as a function of time. 
  Plot parameters are $\Omegatg=0.05\omega$ and $n_0=4$.} 
\end{figure}

The full minimization in third order of expansion, given by the system of
equations (\ref{analiticthird}), (\ref{constr2}) and (\ref{l2}), can be
straightforwardly solved using the exact second-order solution \eqref{secondnoRWA} as an
initial condition for a numerical routine. Fig.~\ref{dynamics}
shows the dynamics over one driving period and illustrates the
striking advantages of polychromatic (PC) driving with $N=10$ frequencies over the MC dynamics. First of all,
because of the approximative identification $H_{\rm eff}\approx H_{\rm tg}$,
the effective dynamics always shows a systematic drift with respect to the
target dynamics. In panels (a) and (b) of Fig.~\ref{dynamics}, the
evolution over a single driving period is compared for short and long
times, respectively. While the MC
evolution with Rabi frequency \eqref{f1MC} deviates significantly from the target evolution after
several driving periods,  the optimal PC dynamics follows the target
rather faithfully. 
As the amplitude of the MC pulse has been chosen in first order, one might wonder if a better performance can be realized with an effective Hamiltonian that includes higher orders.
Such a construction, however, would require a higher driving amplitude and, since the undesired terms in $\Heff^{(2)}$ cannot be set to zero, it results in larger overall deviations with respect to the target dynamics. Thus an improvement of the MC case is not possible through a more accurate treatment.
The lower panel (c) of Fig. \ref{dynamics} shows the second main
advantage, namely
significantly smaller fluctuations of the optimal dynamics around the target dynamics and,
in particular, a considerably lower population of the intermediate
(excited) state.

\begin{figure}
\begin{center}
\begin{overpic}[scale=.5,unit=1mm]
{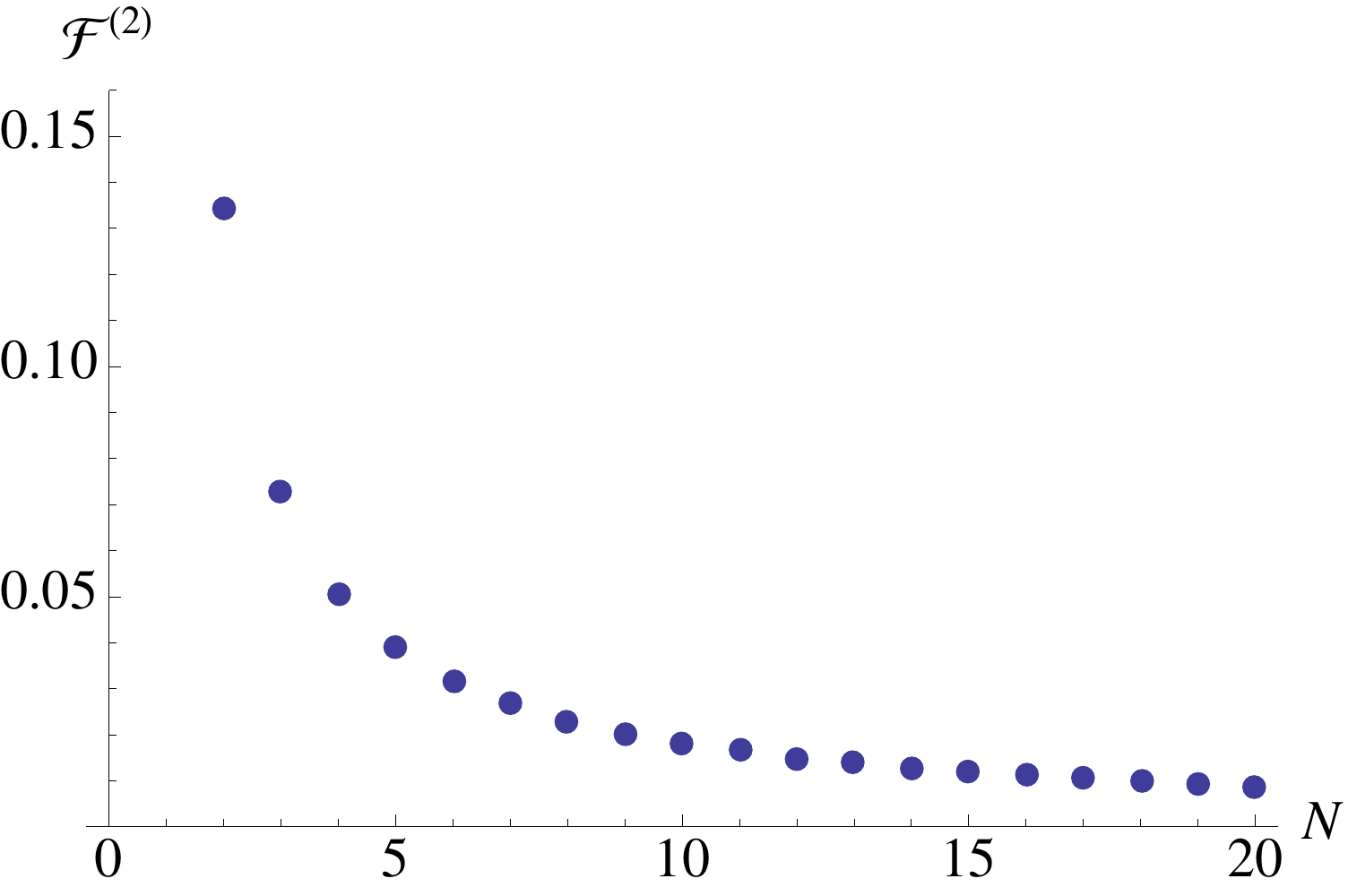}
\put(22,15){\includegraphics[scale=.4]
{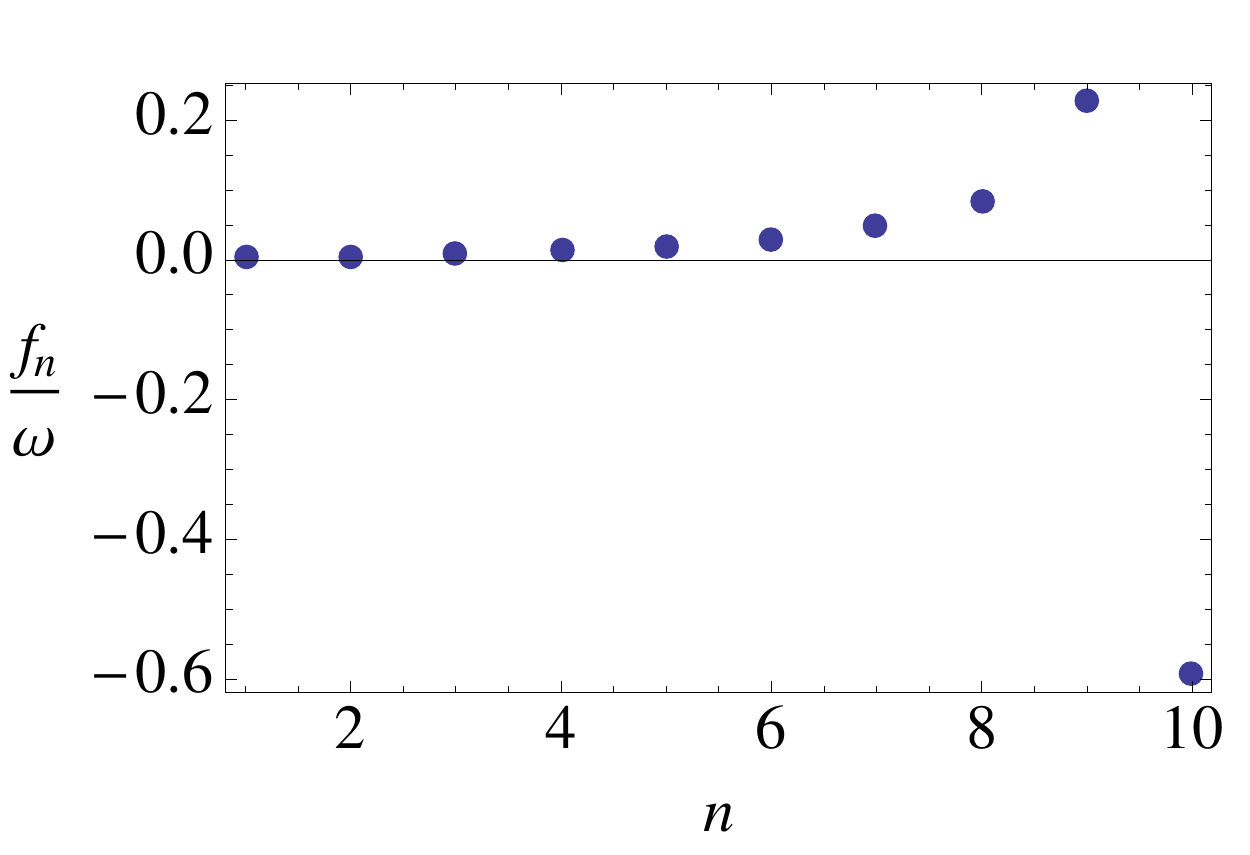}}
\end{overpic}
\end{center}
\caption{Magnitude of fluctuations, quantified by eq.~\eqref{target}, around the
  target unitary dynamics simulated by a third-order optimal pulse, as function of the number of modulation
  Fourier components; plot parameters are $\Omegatg=0.05\omega$ and $n_0=4$. Inset: Fourier
components for $\Nmax=10$. \label{TF}}
\end{figure}

In order to show how much the fluctuations can be suppressed, 
Fig.~\ref{TF} plots the resulting third-order target fidelity 
$\mathcal{F} \approx
\sum_{k=0}^3\mathcal{F}^{\left(k\right)}=\mathcal{F}^{\left(2\right)}$
of  eq.~\eqref{target}, as a function of
the number $N$ of frequency components. Clearly, already a moderate number
of frequency components permits to reduce the fluctuations
dramatically. 
The Fourier components of the 
optimal pulse for $\Nmax=10$ are shown in the inset.  
The component $f_{10}$ of the highest frequency $\Delta= 10\omega$ is
close to the MC solution \eqref{f1MC}; lower frequency components are
phase-shifted by $\pi$ and their magnitude decays rapidly with decreasing frequency.
These results show that modulating the driving with only few
frequencies suffices to simulate the desired target unitary with
significantly higher precision than in the MC case.

\begin{figure}[t]
\includegraphics[scale=0.45]{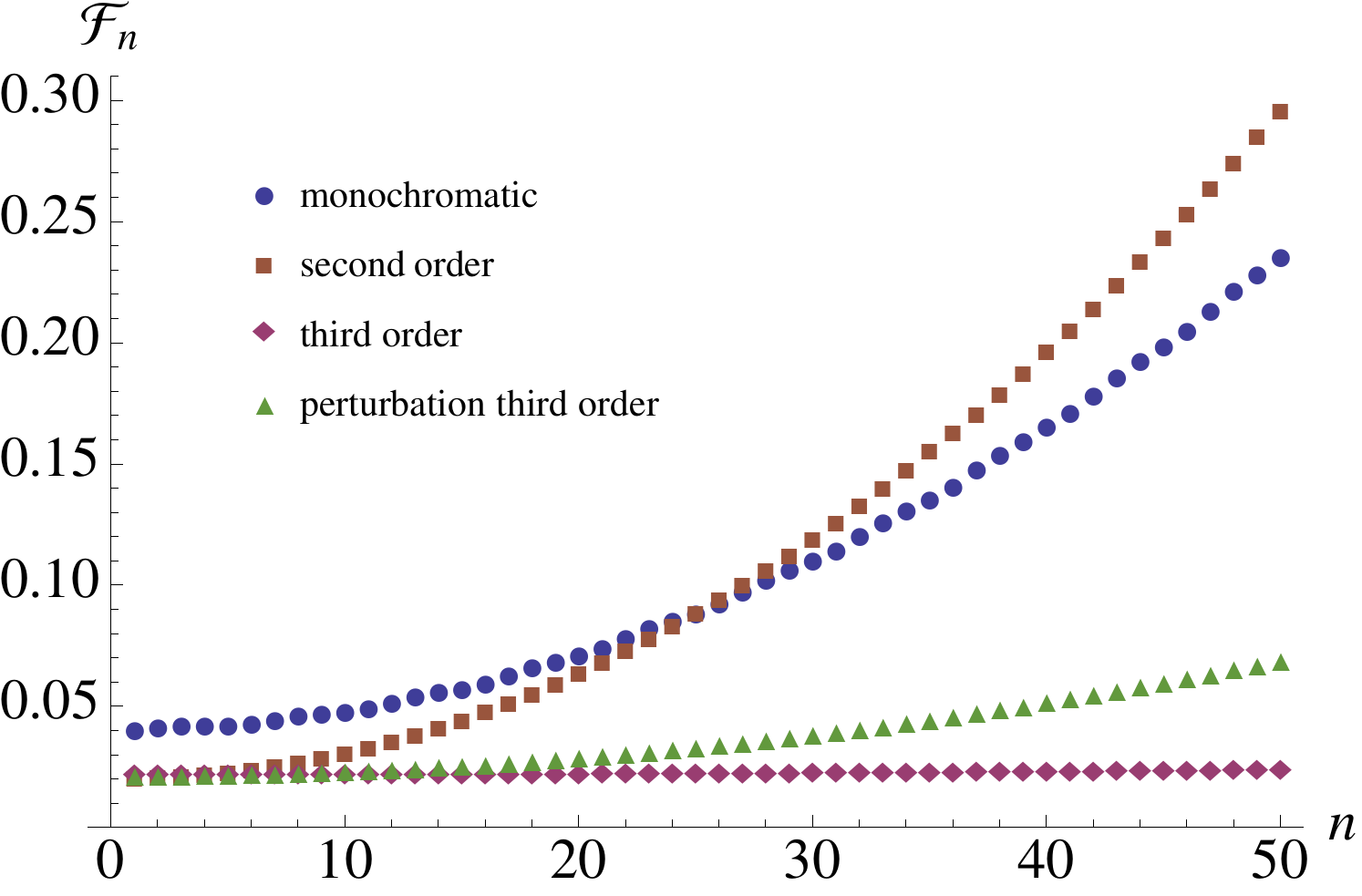}
\caption{\label{deviations}
(color online). Deviations from the target dynamics $\mathcal{F}_n$ in the $n$th driving
period for the MC pulse (blue circles), the second-order optimized
pulse (brown squares) and the third-order optimized pulse (red rhombi)
with $\Nmax=10$. 
The perturbed third order solution [see
Eq. (\ref{perturbations})] with frequency randomness 
$\delta=\delta_\text{max}/4$, averaged over 100 realizations (green
triangles), shows good resilience against experimental uncertainty.  
$\Omegatg=0.05\omega$ and
$n_0=4$.} 
\end{figure}

The long-time deviations observed in Fig. \ref{dynamics}(b) can be
quantitatively measured by the target functional $\mathcal{F}_n$, 
as defined in Eq.~\eqref{targetF},  but integrated over the $n$th driving period. 
In Fig. \ref{deviations}, these fidelities are
shown for the MC pulse and  the second- and third-order PC 
pulses with $\Nmax=10$ as function of $n$. Let us first compare the MC
with the second-order PC pulse. In the first few periods, the optimized solution
indeed yields a better result. However, the deviations with respect to
the target dynamics grow faster in the second order optimized case
than in the MC case. As a consequence, in the long-time regime the MC
driving performs better than the optimized solution calculated in
second order: since the PC pulse contains slower
frequencies than the MC one, the expansion at second order leads to a
worse approximation of the effective Hamiltonian and the deviations
between $U_{\rm eff}$ and $U_{\rm tg}$ accumulate in time and overcome
the difference in fluctuations after a sufficiently long
times. Indeed, we observe that the larger $\Nmax$ is, the later the
crossover occurs, since the fluctuations are smaller and deviations
from the target Hamiltonian need more time to accumulate to the value
of the MC dynamics. 
Nevertheless, the optimized PC pulse can always be 
systematically improved by pushing the calculations to higher order in
the expansion parameter. As seen in Figs.~\ref{dynamics} as well as
\ref{deviations}, the third order optimal pulse
significantly outperforms the MC dynamics in the entire time domain. 

Finally, in order to estimate the robustness of optimal pulses in realistic experimental
setups, we investigate how small perturbations to the
Fourier components affect the performance of the optimal
pulses. Consider perturbations of the form  
\begin{equation}\label{perturbations}
 f_n\rightarrow  \tilde f_n= f_n +\delta{f}
\end{equation}
where $\delta f$ is a random number uniformly distributed in the interval
$[-\delta,\delta]$, which accounts for the
experimental uncertainty in the tuned Fourier
components.  Comparison between the second- and third-order optimal
pulses shows that their largest optimal
Fourier components differ typically by 
$0.01\omega$ (for $\Omegatg=0.05\omega$), which defines a scale
$\delta_\text{max}$ for
the maximum allowed uncertainty. Perturbations with
$\delta=\delta_\text{max}/4$, however, still lead to a good
performance, see Figure \ref{deviations}. Thus, the optimal pulses
appear robust under such perturbations, which indicates a good
experimental viability. 

The control of periodically driven systems by means of pulse shaping presented here opens new perspectives for the optimal simulation of quantum systems.
No increase in intensity as compared to mono-chromatic driving is required and the realization of optimal effective Hamiltonians is robust under perturbations.
The optimal pulses have a rather narrow spectral range, what eases the identification of driving parameters ensuring that no high lying states are excited.
This is of particular importance for large many-body systems, like trapped atomic gases, where un-careful driving easily results in uncontrolled heating.
Since this can be avoided with the present approach, 
it may, for example, be used to enhance or suppress long-range or density-dependent tunneling processes \cite{Verdeny13} in shaken optical lattices.

Financial support by the European Research Council within the project ODYCQUENT is gratefully acknowledged.
F.M. acknowledges
hospitality by the Centre for Quantum Technologies, a Research Centre
of Excellence funded by the Ministry of Education and the
National Research Foundation of Singapore.

\bibliography{mybib}

\begin{thebibliography}{26}%
\makeatletter
\providecommand \@ifxundefined [1]{%
 \@ifx{#1\undefined}
}%
\providecommand \@ifnum [1]{%
 \ifnum #1\expandafter \@firstoftwo
 \else \expandafter \@secondoftwo
 \fi
}%
\providecommand \@ifx [1]{%
 \ifx #1\expandafter \@firstoftwo
 \else \expandafter \@secondoftwo
 \fi
}%
\providecommand \natexlab [1]{#1}%
\providecommand \enquote  [1]{``#1''}%
\providecommand \bibnamefont  [1]{#1}%
\providecommand \bibfnamefont [1]{#1}%
\providecommand \citenamefont [1]{#1}%
\providecommand \href@noop [0]{\@secondoftwo}%
\providecommand \href [0]{\begingroup \@sanitize@url \@href}%
\providecommand \@href[1]{\@@startlink{#1}\@@href}%
\providecommand \@@href[1]{\endgroup#1\@@endlink}%
\providecommand \@sanitize@url [0]{\catcode `\\12\catcode `\$12\catcode
  `\&12\catcode `\#12\catcode `\^12\catcode `\_12\catcode `\%12\relax}%
\providecommand \@@startlink[1]{}%
\providecommand \@@endlink[0]{}%
\providecommand \url  [0]{\begingroup\@sanitize@url \@url }%
\providecommand \@url [1]{\endgroup\@href {#1}{\urlprefix }}%
\providecommand \urlprefix  [0]{URL }%
\providecommand \Eprint [0]{\href }%
\providecommand \doibase [0]{http://dx.doi.org/}%
\providecommand \selectlanguage [0]{\@gobble}%
\providecommand \bibinfo  [0]{\@secondoftwo}%
\providecommand \bibfield  [0]{\@secondoftwo}%
\providecommand \translation [1]{[#1]}%
\providecommand \BibitemOpen [0]{}%
\providecommand \bibitemStop [0]{}%
\providecommand \bibitemNoStop [0]{.\EOS\space}%
\providecommand \EOS [0]{\spacefactor3000\relax}%
\providecommand \BibitemShut  [1]{\csname bibitem#1\endcsname}%
\let\auto@bib@innerbib\@empty
\bibitem [{\citenamefont {Tang}\ \emph {et~al.}(2011)\citenamefont {Tang},
  \citenamefont {Mei},\ and\ \citenamefont {Wen}}]{Tang11}%
  \BibitemOpen
  \bibfield  {author} {\bibinfo {author} {\bibfnamefont {E.}~\bibnamefont
  {Tang}}, \bibinfo {author} {\bibfnamefont {J.-W.}\ \bibnamefont {Mei}}, \
  and\ \bibinfo {author} {\bibfnamefont {X.-G.}\ \bibnamefont {Wen}},\ }\href
  {\doibase 10.1103/PhysRevLett.106.236802} {\bibfield  {journal} {\bibinfo
  {journal} {Phys. Rev. Lett.}\ }\textbf {\bibinfo {volume} {106}},\ \bibinfo
  {pages} {236802} (\bibinfo {year} {2011})}\BibitemShut {NoStop}%
\bibitem [{\citenamefont {Hauke}\ \emph {et~al.}(2012)\citenamefont {Hauke},
  \citenamefont {Tieleman}, \citenamefont {Celi}, \citenamefont
  {\"Olschl\"ager}, \citenamefont {Simonet}, \citenamefont {Struck},
  \citenamefont {Weinberg}, \citenamefont {Windpassinger}, \citenamefont
  {Sengstock}, \citenamefont {Lewenstein},\ and\ \citenamefont
  {Eckardt}}]{Hauke12}%
  \BibitemOpen
  \bibfield  {author} {\bibinfo {author} {\bibfnamefont {P.}~\bibnamefont
  {Hauke}}, \bibinfo {author} {\bibfnamefont {O.}~\bibnamefont {Tieleman}},
  \bibinfo {author} {\bibfnamefont {A.}~\bibnamefont {Celi}}, \bibinfo {author}
  {\bibfnamefont {C.}~\bibnamefont {\"Olschl\"ager}}, \bibinfo {author}
  {\bibfnamefont {J.}~\bibnamefont {Simonet}}, \bibinfo {author} {\bibfnamefont
  {J.}~\bibnamefont {Struck}}, \bibinfo {author} {\bibfnamefont
  {M.}~\bibnamefont {Weinberg}}, \bibinfo {author} {\bibfnamefont
  {P.}~\bibnamefont {Windpassinger}}, \bibinfo {author} {\bibfnamefont
  {K.}~\bibnamefont {Sengstock}}, \bibinfo {author} {\bibfnamefont
  {M.}~\bibnamefont {Lewenstein}}, \ and\ \bibinfo {author} {\bibfnamefont
  {A.}~\bibnamefont {Eckardt}},\ }\href {\doibase
  10.1103/PhysRevLett.109.145301} {\bibfield  {journal} {\bibinfo  {journal}
  {Phys. Rev. Lett.}\ }\textbf {\bibinfo {volume} {109}},\ \bibinfo {pages}
  {145301} (\bibinfo {year} {2012})}\BibitemShut {NoStop}%
\bibitem [{\citenamefont {Banerjee}\ \emph {et~al.}(2013)\citenamefont
  {Banerjee}, \citenamefont {B\"ogli}, \citenamefont {Dalmonte}, \citenamefont
  {Rico}, \citenamefont {Stebler}, \citenamefont {Wiese},\ and\ \citenamefont
  {Zoller}}]{Banerjee13}%
  \BibitemOpen
  \bibfield  {author} {\bibinfo {author} {\bibfnamefont {D.}~\bibnamefont
  {Banerjee}}, \bibinfo {author} {\bibfnamefont {M.}~\bibnamefont {B\"ogli}},
  \bibinfo {author} {\bibfnamefont {M.}~\bibnamefont {Dalmonte}}, \bibinfo
  {author} {\bibfnamefont {E.}~\bibnamefont {Rico}}, \bibinfo {author}
  {\bibfnamefont {P.}~\bibnamefont {Stebler}}, \bibinfo {author} {\bibfnamefont
  {U.-J.}\ \bibnamefont {Wiese}}, \ and\ \bibinfo {author} {\bibfnamefont
  {P.}~\bibnamefont {Zoller}},\ }\href {\doibase
  10.1103/PhysRevLett.110.125303} {\bibfield  {journal} {\bibinfo  {journal}
  {Phys. Rev. Lett.}\ }\textbf {\bibinfo {volume} {110}},\ \bibinfo {pages}
  {125303} (\bibinfo {year} {2013})}\BibitemShut {NoStop}%
\bibitem [{\citenamefont {Goldman}\ \emph {et~al.}(2009)\citenamefont
  {Goldman}, \citenamefont {Kubasiak}, \citenamefont {Bermudez}, \citenamefont
  {Gaspard}, \citenamefont {Lewenstein},\ and\ \citenamefont
  {Martin-Delgado}}]{Goldman09}%
  \BibitemOpen
  \bibfield  {author} {\bibinfo {author} {\bibfnamefont {N.}~\bibnamefont
  {Goldman}}, \bibinfo {author} {\bibfnamefont {A.}~\bibnamefont {Kubasiak}},
  \bibinfo {author} {\bibfnamefont {A.}~\bibnamefont {Bermudez}}, \bibinfo
  {author} {\bibfnamefont {P.}~\bibnamefont {Gaspard}}, \bibinfo {author}
  {\bibfnamefont {M.}~\bibnamefont {Lewenstein}}, \ and\ \bibinfo {author}
  {\bibfnamefont {M.~A.}\ \bibnamefont {Martin-Delgado}},\ }\href {\doibase
  10.1103/PhysRevLett.103.035301} {\bibfield  {journal} {\bibinfo  {journal}
  {Phys. Rev. Lett.}\ }\textbf {\bibinfo {volume} {103}},\ \bibinfo {pages}
  {035301} (\bibinfo {year} {2009})}\BibitemShut {NoStop}%
\bibitem [{\citenamefont {Franco-Villafa\~ne}\ \emph
  {et~al.}(2013)\citenamefont {Franco-Villafa\~ne}, \citenamefont {Sadurn\'\i},
  \citenamefont {Barkhofen}, \citenamefont {Kuhl}, \citenamefont
  {Mortessagne},\ and\ \citenamefont {Seligman}}]{FV13}%
  \BibitemOpen
  \bibfield  {author} {\bibinfo {author} {\bibfnamefont {J.~A.}\ \bibnamefont
  {Franco-Villafa\~ne}}, \bibinfo {author} {\bibfnamefont {E.}~\bibnamefont
  {Sadurn\'\i}}, \bibinfo {author} {\bibfnamefont {S.}~\bibnamefont
  {Barkhofen}}, \bibinfo {author} {\bibfnamefont {U.}~\bibnamefont {Kuhl}},
  \bibinfo {author} {\bibfnamefont {F.}~\bibnamefont {Mortessagne}}, \ and\
  \bibinfo {author} {\bibfnamefont {T.~H.}\ \bibnamefont {Seligman}},\ }\href
  {\doibase 10.1103/PhysRevLett.111.170405} {\bibfield  {journal} {\bibinfo
  {journal} {Phys. Rev. Lett.}\ }\textbf {\bibinfo {volume} {111}},\ \bibinfo
  {pages} {170405} (\bibinfo {year} {2013})}\BibitemShut {NoStop}%
\bibitem [{\citenamefont {Cheuk}\ \emph {et~al.}(2012)\citenamefont {Cheuk},
  \citenamefont {Sommer}, \citenamefont {Hadzibabic}, \citenamefont {Yefsah},
  \citenamefont {Bakr},\ and\ \citenamefont {Zwierlein}}]{Cheuk12}%
  \BibitemOpen
  \bibfield  {author} {\bibinfo {author} {\bibfnamefont {L.~W.}\ \bibnamefont
  {Cheuk}}, \bibinfo {author} {\bibfnamefont {A.~T.}\ \bibnamefont {Sommer}},
  \bibinfo {author} {\bibfnamefont {Z.}~\bibnamefont {Hadzibabic}}, \bibinfo
  {author} {\bibfnamefont {T.}~\bibnamefont {Yefsah}}, \bibinfo {author}
  {\bibfnamefont {W.~S.}\ \bibnamefont {Bakr}}, \ and\ \bibinfo {author}
  {\bibfnamefont {M.~W.}\ \bibnamefont {Zwierlein}},\ }\href {\doibase
  10.1103/PhysRevLett.109.095302} {\bibfield  {journal} {\bibinfo  {journal}
  {Phys. Rev. Lett.}\ }\textbf {\bibinfo {volume} {109}},\ \bibinfo {pages}
  {095302} (\bibinfo {year} {2012})}\BibitemShut {NoStop}%
\bibitem [{\citenamefont {Wang}\ \emph {et~al.}(2012)\citenamefont {Wang},
  \citenamefont {Yu}, \citenamefont {Fu}, \citenamefont {Miao}, \citenamefont
  {Huang}, \citenamefont {Chai}, \citenamefont {Zhai},\ and\ \citenamefont
  {Zhang}}]{Wang12}%
  \BibitemOpen
  \bibfield  {author} {\bibinfo {author} {\bibfnamefont {P.}~\bibnamefont
  {Wang}}, \bibinfo {author} {\bibfnamefont {Z.-Q.}\ \bibnamefont {Yu}},
  \bibinfo {author} {\bibfnamefont {Z.}~\bibnamefont {Fu}}, \bibinfo {author}
  {\bibfnamefont {J.}~\bibnamefont {Miao}}, \bibinfo {author} {\bibfnamefont
  {L.}~\bibnamefont {Huang}}, \bibinfo {author} {\bibfnamefont
  {S.}~\bibnamefont {Chai}}, \bibinfo {author} {\bibfnamefont {H.}~\bibnamefont
  {Zhai}}, \ and\ \bibinfo {author} {\bibfnamefont {J.}~\bibnamefont {Zhang}},\
  }\href {\doibase 10.1103/PhysRevLett.109.095301} {\bibfield  {journal}
  {\bibinfo  {journal} {Phys. Rev. Lett.}\ }\textbf {\bibinfo {volume} {109}},\
  \bibinfo {pages} {095301} (\bibinfo {year} {2012})}\BibitemShut {NoStop}%
\bibitem [{\citenamefont {Miyake}\ \emph {et~al.}(2013)\citenamefont {Miyake},
  \citenamefont {Siviloglou}, \citenamefont {Kennedy}, \citenamefont {Burton},\
  and\ \citenamefont {Ketterle}}]{Miyake13}%
  \BibitemOpen
  \bibfield  {author} {\bibinfo {author} {\bibfnamefont {H.}~\bibnamefont
  {Miyake}}, \bibinfo {author} {\bibfnamefont {G.~A.}\ \bibnamefont
  {Siviloglou}}, \bibinfo {author} {\bibfnamefont {C.~J.}\ \bibnamefont
  {Kennedy}}, \bibinfo {author} {\bibfnamefont {W.~C.}\ \bibnamefont {Burton}},
  \ and\ \bibinfo {author} {\bibfnamefont {W.}~\bibnamefont {Ketterle}},\
  }\href {\doibase 10.1103/PhysRevLett.111.185302} {\bibfield  {journal}
  {\bibinfo  {journal} {Phys. Rev. Lett.}\ }\textbf {\bibinfo {volume} {111}},\
  \bibinfo {pages} {185302} (\bibinfo {year} {2013})}\BibitemShut {NoStop}%
\bibitem [{\citenamefont {Aidelsburger}\ \emph {et~al.}(2013)\citenamefont
  {Aidelsburger}, \citenamefont {Atala}, \citenamefont {Lohse}, \citenamefont
  {Barreiro}, \citenamefont {Paredes},\ and\ \citenamefont
  {Bloch}}]{Aidelsburger13}%
  \BibitemOpen
  \bibfield  {author} {\bibinfo {author} {\bibfnamefont {M.}~\bibnamefont
  {Aidelsburger}}, \bibinfo {author} {\bibfnamefont {M.}~\bibnamefont {Atala}},
  \bibinfo {author} {\bibfnamefont {M.}~\bibnamefont {Lohse}}, \bibinfo
  {author} {\bibfnamefont {J.~T.}\ \bibnamefont {Barreiro}}, \bibinfo {author}
  {\bibfnamefont {B.}~\bibnamefont {Paredes}}, \ and\ \bibinfo {author}
  {\bibfnamefont {I.}~\bibnamefont {Bloch}},\ }\href {\doibase
  10.1103/PhysRevLett.111.185301} {\bibfield  {journal} {\bibinfo  {journal}
  {Phys. Rev. Lett.}\ }\textbf {\bibinfo {volume} {111}},\ \bibinfo {pages}
  {185301} (\bibinfo {year} {2013})}\BibitemShut {NoStop}%
\bibitem [{\citenamefont {Eckardt}\ \emph {et~al.}(2005)\citenamefont
  {Eckardt}, \citenamefont {Weiss},\ and\ \citenamefont
  {Holthaus}}]{Eckardt05}%
  \BibitemOpen
  \bibfield  {author} {\bibinfo {author} {\bibfnamefont {A.}~\bibnamefont
  {Eckardt}}, \bibinfo {author} {\bibfnamefont {C.}~\bibnamefont {Weiss}}, \
  and\ \bibinfo {author} {\bibfnamefont {M.}~\bibnamefont {Holthaus}},\ }\href
  {\doibase 10.1103/PhysRevLett.95.260404} {\bibfield  {journal} {\bibinfo
  {journal} {Phys. Rev. Lett.}\ }\textbf {\bibinfo {volume} {95}},\ \bibinfo
  {pages} {260404} (\bibinfo {year} {2005})}\BibitemShut {NoStop}%
\bibitem [{\citenamefont {Lignier}\ \emph {et~al.}(2007)\citenamefont
  {Lignier}, \citenamefont {Sias}, \citenamefont {Ciampini}, \citenamefont
  {Singh}, \citenamefont {Zenesini}, \citenamefont {Morsch},\ and\
  \citenamefont {Arimondo}}]{Lignier07}%
  \BibitemOpen
  \bibfield  {author} {\bibinfo {author} {\bibfnamefont {H.}~\bibnamefont
  {Lignier}}, \bibinfo {author} {\bibfnamefont {C.}~\bibnamefont {Sias}},
  \bibinfo {author} {\bibfnamefont {D.}~\bibnamefont {Ciampini}}, \bibinfo
  {author} {\bibfnamefont {Y.}~\bibnamefont {Singh}}, \bibinfo {author}
  {\bibfnamefont {A.}~\bibnamefont {Zenesini}}, \bibinfo {author}
  {\bibfnamefont {O.}~\bibnamefont {Morsch}}, \ and\ \bibinfo {author}
  {\bibfnamefont {E.}~\bibnamefont {Arimondo}},\ }\href {\doibase
  10.1103/PhysRevLett.99.220403} {\bibfield  {journal} {\bibinfo  {journal}
  {Phys. Rev. Lett.}\ }\textbf {\bibinfo {volume} {99}},\ \bibinfo {pages}
  {220403} (\bibinfo {year} {2007})}\BibitemShut {NoStop}%
\bibitem [{\citenamefont {Struck}\ \emph {et~al.}(2012)\citenamefont {Struck},
  \citenamefont {{\"O}lschl{\"a}ger}, \citenamefont {Weinberg}, \citenamefont
  {Hauke}, \citenamefont {Simonet}, \citenamefont {Eckardt}, \citenamefont
  {Lewenstein}, \citenamefont {Sengstock},\ and\ \citenamefont
  {Windpassinger}}]{Struck12}%
  \BibitemOpen
  \bibfield  {author} {\bibinfo {author} {\bibfnamefont {J.}~\bibnamefont
  {Struck}}, \bibinfo {author} {\bibfnamefont {C.}~\bibnamefont
  {{\"O}lschl{\"a}ger}}, \bibinfo {author} {\bibfnamefont {M.}~\bibnamefont
  {Weinberg}}, \bibinfo {author} {\bibfnamefont {P.}~\bibnamefont {Hauke}},
  \bibinfo {author} {\bibfnamefont {J.}~\bibnamefont {Simonet}}, \bibinfo
  {author} {\bibfnamefont {A.}~\bibnamefont {Eckardt}}, \bibinfo {author}
  {\bibfnamefont {M.}~\bibnamefont {Lewenstein}}, \bibinfo {author}
  {\bibfnamefont {K.}~\bibnamefont {Sengstock}}, \ and\ \bibinfo {author}
  {\bibfnamefont {P.}~\bibnamefont {Windpassinger}},\ }\href {\doibase
  10.1103/PhysRevLett.108.225304} {\bibfield  {journal} {\bibinfo  {journal}
  {Phys. Rev. Lett.}\ }\textbf {\bibinfo {volume} {108}},\ \bibinfo {pages}
  {225304} (\bibinfo {year} {2012})}\BibitemShut {NoStop}%
\bibitem [{\citenamefont {Rapp}\ \emph {et~al.}(2012)\citenamefont {Rapp},
  \citenamefont {Deng},\ and\ \citenamefont {Santos}}]{Rapp12}%
  \BibitemOpen
  \bibfield  {author} {\bibinfo {author} {\bibfnamefont {A.}~\bibnamefont
  {Rapp}}, \bibinfo {author} {\bibfnamefont {X.}~\bibnamefont {Deng}}, \ and\
  \bibinfo {author} {\bibfnamefont {L.}~\bibnamefont {Santos}},\ }\href
  {\doibase 10.1103/PhysRevLett.109.203005} {\bibfield  {journal} {\bibinfo
  {journal} {Phys. Rev. Lett.}\ }\textbf {\bibinfo {volume} {109}},\ \bibinfo
  {pages} {203005} (\bibinfo {year} {2012})}\BibitemShut {NoStop}%
\bibitem [{\citenamefont {Iadecola}\ \emph {et~al.}(2013)\citenamefont
  {Iadecola}, \citenamefont {Campbell}, \citenamefont {Chamon}, \citenamefont
  {Hou}, \citenamefont {Jackiw}, \citenamefont {Pi},\ and\ \citenamefont
  {Kusminskiy}}]{Iadecola13}%
  \BibitemOpen
  \bibfield  {author} {\bibinfo {author} {\bibfnamefont {T.}~\bibnamefont
  {Iadecola}}, \bibinfo {author} {\bibfnamefont {D.}~\bibnamefont {Campbell}},
  \bibinfo {author} {\bibfnamefont {C.}~\bibnamefont {Chamon}}, \bibinfo
  {author} {\bibfnamefont {C.-Y.}\ \bibnamefont {Hou}}, \bibinfo {author}
  {\bibfnamefont {R.}~\bibnamefont {Jackiw}}, \bibinfo {author} {\bibfnamefont
  {S.-Y.}\ \bibnamefont {Pi}}, \ and\ \bibinfo {author} {\bibfnamefont {S.~V.}\
  \bibnamefont {Kusminskiy}},\ }\href {\doibase 10.1103/PhysRevLett.110.176603}
  {\bibfield  {journal} {\bibinfo  {journal} {Phys. Rev. Lett.}\ }\textbf
  {\bibinfo {volume} {110}},\ \bibinfo {pages} {176603} (\bibinfo {year}
  {2013})}\BibitemShut {NoStop}%
\bibitem [{\citenamefont {Braun}\ \emph {et~al.}(2013)\citenamefont {Braun},
  \citenamefont {Ronzheimer}, \citenamefont {Schreiber}, \citenamefont
  {Hodgman}, \citenamefont {Rom}, \citenamefont {Bloch},\ and\ \citenamefont
  {Schneider}}]{Braun13}%
  \BibitemOpen
  \bibfield  {author} {\bibinfo {author} {\bibfnamefont {S.}~\bibnamefont
  {Braun}}, \bibinfo {author} {\bibfnamefont {J.~P.}\ \bibnamefont
  {Ronzheimer}}, \bibinfo {author} {\bibfnamefont {M.}~\bibnamefont
  {Schreiber}}, \bibinfo {author} {\bibfnamefont {S.~S.}\ \bibnamefont
  {Hodgman}}, \bibinfo {author} {\bibfnamefont {T.}~\bibnamefont {Rom}},
  \bibinfo {author} {\bibfnamefont {I.}~\bibnamefont {Bloch}}, \ and\ \bibinfo
  {author} {\bibfnamefont {U.}~\bibnamefont {Schneider}},\ }\href {\doibase
  10.1126/science.1227831} {\bibfield  {journal} {\bibinfo  {journal}
  {Science}\ }\textbf {\bibinfo {volume} {339}},\ \bibinfo {pages} {52}
  (\bibinfo {year} {2013})}\BibitemShut {NoStop}%
\bibitem [{\citenamefont {Struck}\ \emph {et~al.}(2013)\citenamefont {Struck},
  \citenamefont {Weinberg}, \citenamefont {Olschlager}, \citenamefont
  {Windpassinger}, \citenamefont {Simonet}, \citenamefont {Sengstock},
  \citenamefont {Hoppner}, \citenamefont {Hauke}, \citenamefont {Eckardt},
  \citenamefont {Lewenstein},\ and\ \citenamefont {Mathey}}]{Struck13}%
  \BibitemOpen
  \bibfield  {author} {\bibinfo {author} {\bibfnamefont {J.}~\bibnamefont
  {Struck}}, \bibinfo {author} {\bibfnamefont {M.}~\bibnamefont {Weinberg}},
  \bibinfo {author} {\bibfnamefont {C.}~\bibnamefont {Olschlager}}, \bibinfo
  {author} {\bibfnamefont {P.}~\bibnamefont {Windpassinger}}, \bibinfo {author}
  {\bibfnamefont {J.}~\bibnamefont {Simonet}}, \bibinfo {author} {\bibfnamefont
  {K.}~\bibnamefont {Sengstock}}, \bibinfo {author} {\bibfnamefont
  {R.}~\bibnamefont {Hoppner}}, \bibinfo {author} {\bibfnamefont
  {P.}~\bibnamefont {Hauke}}, \bibinfo {author} {\bibfnamefont
  {A.}~\bibnamefont {Eckardt}}, \bibinfo {author} {\bibfnamefont
  {M.}~\bibnamefont {Lewenstein}}, \ and\ \bibinfo {author} {\bibfnamefont
  {L.}~\bibnamefont {Mathey}},\ }\href {http://dx.doi.org/10.1038/nphys2750}
  {\bibfield  {journal} {\bibinfo  {journal} {Nat Phys}\ }\textbf {\bibinfo
  {volume} {9}},\ \bibinfo {pages} {738} (\bibinfo {year} {2013})}\BibitemShut
  {NoStop}%
\bibitem [{\citenamefont {D'Alessandro}(2007)}]{Alessandro07}%
  \BibitemOpen
  \bibfield  {author} {\bibinfo {author} {\bibfnamefont {D.}~\bibnamefont
  {D'Alessandro}},\ }\href@noop {} {\emph {\bibinfo {title} {Introduction to
  Quantum Control and Dynamics}}}\ (\bibinfo  {publisher} {CRC, Boca Raton},\
  \bibinfo {year} {2007})\BibitemShut {NoStop}%
\bibitem [{\citenamefont {Krotov}(1996)}]{Krotov96}%
  \BibitemOpen
  \bibfield  {author} {\bibinfo {author} {\bibfnamefont {V.~F.}\ \bibnamefont
  {Krotov}},\ }\href@noop {} {\emph {\bibinfo {title} {Global Methods in
  Optimal Control Theory}}}\ (\bibinfo  {publisher} {Marcel Dekker, New York},\
  \bibinfo {year} {1996})\BibitemShut {NoStop}%
\bibitem [{\citenamefont {Khaneja}\ \emph {et~al.}(2005)\citenamefont
  {Khaneja}, \citenamefont {Reiss}, \citenamefont {Kehlet}, \citenamefont
  {Schulte-Herbr{\"u}ggen},\ and\ \citenamefont {Glaser}}]{Khaneja05}%
  \BibitemOpen
  \bibfield  {author} {\bibinfo {author} {\bibfnamefont {N.}~\bibnamefont
  {Khaneja}}, \bibinfo {author} {\bibfnamefont {T.}~\bibnamefont {Reiss}},
  \bibinfo {author} {\bibfnamefont {C.}~\bibnamefont {Kehlet}}, \bibinfo
  {author} {\bibfnamefont {T.}~\bibnamefont {Schulte-Herbr{\"u}ggen}}, \ and\
  \bibinfo {author} {\bibfnamefont {S.~J.}\ \bibnamefont {Glaser}},\ }\href
  {\doibase http://dx.doi.org/10.1016/j.jmr.2004.11.004} {\bibfield  {journal}
  {\bibinfo  {journal} {J. Magn. Res.}\ }\textbf {\bibinfo {volume} {172}},\
  \bibinfo {pages} {296 } (\bibinfo {year} {2005})}\BibitemShut {NoStop}%
\bibitem [{\citenamefont {Reich}\ \emph {et~al.}(2013)\citenamefont {Reich},
  \citenamefont {Gualdi},\ and\ \citenamefont {Koch}}]{Reich13}%
  \BibitemOpen
  \bibfield  {author} {\bibinfo {author} {\bibfnamefont {D.~M.}\ \bibnamefont
  {Reich}}, \bibinfo {author} {\bibfnamefont {G.}~\bibnamefont {Gualdi}}, \
  and\ \bibinfo {author} {\bibfnamefont {C.~P.}\ \bibnamefont {Koch}},\ }\href
  {\doibase 10.1103/PhysRevLett.111.200401} {\bibfield  {journal} {\bibinfo
  {journal} {Phys. Rev. Lett.}\ }\textbf {\bibinfo {volume} {111}},\ \bibinfo
  {pages} {200401} (\bibinfo {year} {2013})}\BibitemShut {NoStop}%
\bibitem [{\citenamefont {Bartels}\ and\ \citenamefont
  {Mintert}(2013)}]{Bartels13}%
  \BibitemOpen
  \bibfield  {author} {\bibinfo {author} {\bibfnamefont {B.}~\bibnamefont
  {Bartels}}\ and\ \bibinfo {author} {\bibfnamefont {F.}~\bibnamefont
  {Mintert}},\ }\href {\doibase 10.1103/PhysRevA.88.052315} {\bibfield
  {journal} {\bibinfo  {journal} {Phys. Rev. A}\ }\textbf {\bibinfo {volume}
  {88}},\ \bibinfo {pages} {052315} (\bibinfo {year} {2013})}\BibitemShut
  {NoStop}%
\bibitem [{\citenamefont {Floquet}(1883)}]{Floquet83}%
  \BibitemOpen
  \bibfield  {author} {\bibinfo {author} {\bibfnamefont {G.}~\bibnamefont
  {Floquet}},\ }\href@noop {} {\bibfield  {journal} {\bibinfo  {journal} {Ann.
  {\'E}cole Norm. Sup.}\ }\textbf {\bibinfo {volume} {12}},\ \bibinfo {pages}
  {47} (\bibinfo {year} {1883})}\BibitemShut {NoStop}%
\bibitem [{\citenamefont {Magnus}(1954)}]{Magnus54}%
  \BibitemOpen
  \bibfield  {author} {\bibinfo {author} {\bibfnamefont {W.}~\bibnamefont
  {Magnus}},\ }\href {\doibase 10.1002/cpa.3160070404} {\bibfield  {journal}
  {\bibinfo  {journal} {Comm. P. and App. Math.}\ }\textbf {\bibinfo {volume}
  {7}},\ \bibinfo {pages} {649} (\bibinfo {year} {1954})}\BibitemShut {NoStop}%
\bibitem [{\citenamefont {Blanes}\ \emph {et~al.}(2009)\citenamefont {Blanes},
  \citenamefont {Casas}, \citenamefont {Oteo},\ and\ \citenamefont
  {Ros}}]{Blanes09}%
  \BibitemOpen
  \bibfield  {author} {\bibinfo {author} {\bibfnamefont {S.}~\bibnamefont
  {Blanes}}, \bibinfo {author} {\bibfnamefont {F.}~\bibnamefont {Casas}},
  \bibinfo {author} {\bibfnamefont {J.}~\bibnamefont {Oteo}}, \ and\ \bibinfo
  {author} {\bibfnamefont {J.}~\bibnamefont {Ros}},\ }\href {\doibase
  http://dx.doi.org/10.1016/j.physrep.2008.11.001} {\bibfield  {journal}
  {\bibinfo  {journal} {Phys. Rep.}\ }\textbf {\bibinfo {volume} {470}},\
  \bibinfo {pages} {151 } (\bibinfo {year} {2009})}\BibitemShut {NoStop}%
\bibitem [{Note1()}]{Note1}%
  \BibitemOpen
  \bibinfo {note} {These terms also generate light-shift displacements of the
  excited and ground states which do not influence the target dynamics between
  the two ground states.}\BibitemShut {Stop}%
\bibitem [{\citenamefont {Verdeny}\ \emph {et~al.}(2013)\citenamefont
  {Verdeny}, \citenamefont {Mielke},\ and\ \citenamefont
  {Mintert}}]{Verdeny13}%
  \BibitemOpen
  \bibfield  {author} {\bibinfo {author} {\bibfnamefont {A.}~\bibnamefont
  {Verdeny}}, \bibinfo {author} {\bibfnamefont {A.}~\bibnamefont {Mielke}}, \
  and\ \bibinfo {author} {\bibfnamefont {F.}~\bibnamefont {Mintert}},\ }\href
  {\doibase 10.1103/PhysRevLett.111.175301} {\bibfield  {journal} {\bibinfo
  {journal} {Phys. Rev. Lett.}\ }\textbf {\bibinfo {volume} {111}},\ \bibinfo
  {pages} {175301} (\bibinfo {year} {2013})}\BibitemShut {NoStop}%
\end{thebibliography}%

\end{document}